\documentclass[sn-mathphys]{sn-jnl}

\theoremstyle{thmstyleone}%
\theoremstyle{thmstyletwo}%

\theoremstyle{thmstylethree}%

\usepackage{color}
\newcommand{\blue}{\color{blue}}

\usepackage{graphicx}
\usepackage{amssymb,bm}
\usepackage{amsmath,amsthm}
\usepackage{epstopdf}
\usepackage{hyperref}
\hypersetup{
    colorlinks=true,
    linkcolor=blue,
    filecolor=magenta,
    urlcolor=cyan,
    citecolor=magenta
}
\newcommand{\be}{\begin{equation}} 
\newcommand{\ee}{\end{equation}}

\begin{document}

\title[]{Forces parallel to particle trajectories in relativistic gravity}


\author[1]{\fnm{Valerio} \sur{Faraoni}}\email{vfaraoni@ubishops.ca}
\author[2]{\fnm{Luca} \sur{Valsan}}\email{luca.valsan@mail.mcgill.ca}
\author[1]{\fnm{Uri} \sur{Gorman}}\email{ugorman24@ubishops.ca}

\affil[1]{\orgdiv{Department of Physics \& Astronomy}, \orgname{Bishop's 
University}, \orgaddress{\street{2600 College Street}, \city{Sherbrooke}, 
\postcode{J1M~1Z7}, \state{Qu\'ebec}, \country{Canada}}}

\affil[2]{\orgdiv{Department of Physics}, \orgname{McGill  
University}, \orgaddress{\street{3600 rue University}, \city{Montr\'eal}, 
\postcode{H3A~2T8}, \state{Qu\'ebec}, \country{Canada}}}

\abstract{Forces parallel to particle trajectories occur in physically 
meaningful situations, including relativistic cosmology and Einstein frame 
scalar-tensor gravity. These situations have Newtonian analogues that we 
discuss to provide intuition about the underlying physics.
 }




\maketitle

\section{Introduction}
\label{sec:1}
\setcounter{equation}{0}

In general relativity (GR), massive test particles follow timelike 
geodesic curves, a result deduced from the hydrodynamical equations for a 
dust (e.g., \cite{Waldbook}). In a geometry described by the spacetime 
metric\footnote{We follow the notation of Ref.~\cite{Waldbook}.}   
$g_{ab}$ and covariant derivative operator $\nabla_a$, a geodesic curve 
with four-tangent $u^a$ (normalized to $u_c u^c=-1$) 
can be affinely parametrized and described by 
$u^a\nabla_a u^b=0$, or non-affinely parametrized and satisfying instead 
\be
u^a\nabla_a u^b= \alpha \, u^b \label{non-affine}
\ee
or, in coordinates,
\be
\frac{d^2 x^a}{d \lambda^2} + \Gamma^a_{bc} \, \frac{dx^b}{d\lambda} \, 
\frac{dx^c}{d\lambda} =\alpha \, \frac{dx^a}{d\lambda} \,,
\ee
where $\lambda$ is a non-affine parameter along the curve and $\alpha$ is 
a function of the spacetime position. The right-hand 
side of Eq.~(\ref{non-affine}) has the appearance of a force (per unit 
mass) parallel to the 
particle's  trajectory but,  
since its nature is purely gravitational, it can be eliminated by 
reparametrizing the curve with  a new affine parameter (usually the 
proper time $\tau$), then the geodesic equation reduces to the 
homogeneous one $u^a\nabla_a u^b=0$.  

The question 
arises: what about non-gravitational forces that are parallel to the 
trajectory? A massive test particle subject to such a force obeys 
\be
\frac{d^2 x^a}{d \tau^2} + \Gamma^a_{bc} \, \frac{dx^b}{d\tau} \, 
\frac{dx^c}{d\tau} =\alpha \, \frac{dx^a}{d\tau} \,,
\ee
where $\tau$ is the proper time, which now is {\em not} an affine 
parameter. The mathematical procedure to 
reparametrize the curve does not care about the nature of the apparent 
force, hence these ``real'' parallel  forces can always be eliminated as 
well, at the price of using a parameter different from the proper time.  
While, from the  mathematical point of view, all 
parametrizations are equivalent and a curve is just an equivalence class 
of parametrizations, different parameters may have different physical 
meanings, for example the proper time is certainly  a physically 
preferred 
parameter while,  for all practical purposes, in cosmology one uses the 
time 
of observers comoving with the cosmic substratum, or the conformal time. 
Nevertheless, the ease of eliminating forces parallel to particle 
trajectories makes these trajectories special---they have been dubbed 
``quasi-geodesics'' \cite{Faraoni:2020ejh}. The time 
reparametrization is usually related to the use of different coordinate 
systems in  black hole physics and cosmology 
\cite{Gullstrand1912, Painleve1921, Eddington:1924pmh,Finkelstein1958,
Eriksen:1995ws, Martel:2000rn, Pascu:2012yu,Fromholz:2013hka,
Volovik:2022cay, Faraoni:2020ehi,Vachon:2021bya,Faraoni:2022coe, 
PhysRevD.29.198, PhysRevD.17.2552}.  

Are ``real'' (i.e., non-gravitational) forces parallel to the trajectories 
of massive test particles relevant at all? While they are not at all 
general in the theory, there are important physical situations where such 
forces appear, including Friedmann-Lema\^itre-Robertson-Walker (FLRW)  
cosmology, self-interacting dark matter scenarios, and 
Einstein 
frame scalar-tensor gravity \cite{Brans:1961sx}, and they are not mere 
curiosities. In the following sections we discuss these forces, aiming to 
provide physical intuition by means of analogies with Newtonian mechanics. 
In the Einstein frame, a parallel force has (correctly) been interpreted 
in early literature as a manifestation of the fact that the particle mass 
varies along the trajectory \cite{Dicke:1961gz}.  This situation exhibits 
a Newtonian analogy with a particle of varying mass $m(t)$, and its study 
brings to light the fact that forces parallel to the particle velocity can 
be eliminated in Newtonian physics. The elimination of forces as a formal 
trick is a recurrent theme in Newtonian physics 
\cite{Landau:1975pou,Goldstein}, but it is better appreciated in the wider 
picture including GR and scalar-tensor gravity, adding a pedagogical 
flavour to the discussion. Of course, in Newtonian mechanics there is an 
absolute time $t$ and it does not make sense physically to change this 
parameter along particle trajectories: trajectory reparametrization is a 
purely formal trick, but it appears in standard textbooks 
\cite{Landau:1975pou,Goldstein}, and it may even be convenient for solving 
the equations of motion explicitly \cite{Kobe1993}. For example, in 
non-relativistic Lagrangian mechanics the Jacobi form of the least action 
(or Maupertuis)  principle formally eliminates a time-independent 
three-force acting on a particle by introducing a fictitious curved 
geometry such that the particle follows one of its geodesics 
\cite{Landau:1975pou,Goldstein}. Thus, the motion of the representative 
point of the particle in configuration space is reduced to a geodesic flow 
in a suitable Riemannian metric $\tilde{g}_{ij}$ (in general, this 
reduction is non-trivial 
\cite{Kolmogorov1954,Abe1984,Glass1977,Lynch1985}). The three-force can 
only be eliminated along a single curve, not across an entire region of 
space and $\tilde{g}_{ij}$ has no physical meaning. This procedure 
involves the introduction of a new parameter along the curve, replacing 
the absolute Newtonian time originally used as a parameter.

Forces {\it orthogonal} to the trajectories of massive test particles have 
already been studied by Price \cite{Price:2004ns}, but their usefulness is 
limited to stationary spacetimes. These forces perform zero work, in 
analogy with the magnetic force acting on an electric charge in three 
dimensions.

A four-force (per unit mass) parallel to the timelike trajectory of a 
massive particle has the form $F^a= \alpha \, u^a $, the particle's 
equation of motion $ u^a\nabla_a u^b= \alpha \, u^b $ is the non-affinely 
parametrized geodesic equation, and the proper time $\tau$ is not an 
affine parameter. It is always possible to eliminate this parallel 
four-force by finding an affine parametrization.

Although the relativistic  situations that we discuss should in 
principle be well-known,  it seems that the literature lacks simple 
explanations where  Newtonian analogies facilitate the physical 
undertanding. Moreover, these 
separate occurrences of parallel forces are never discussed together.  
FLRW cosmology with a non-zero pressure is the basis 
of the standard $\Lambda$-Cold Dark Matter model of the universe, yet 
parallel forces are not mentioned explicitly in cosmology or GR textbooks, 
which only employ comoving or conformal time. In the following we provide 
an explicit and unified discussion of this subject.

\section{Parallel forces in cosmology}
\label{sec:2}
\setcounter{equation}{0}

Consider a spatially homogeneous and isotropic FLRW universe with line 
element
\be
ds^2= -dt^2 +a^2(t) \left[ \frac{dr^2}{1-kr^2} + r^2 \left( 
d\vartheta^2 + \sin^2\vartheta \, d\varphi^2 \right)\right]
=g_{ab} dx^a dx^b 
\ee
in comoving polar coordinates $\left(t,r,\vartheta, \varphi \right)$.   
Assume that the matter source of this FLRW 
geometry is a perfect fluid with stress-energy tensor
\be
T_{ab} =\left( P+\rho\right) u_au_b +P g_{ab} \,,
\ee
where $u^a$ is the fluid 4-velocity, $\rho$ is the energy density, and $P$ 
is the isotropic pressure, which is taken to be non-zero and 
non-constant.\footnote{This assumption excludes the case of a   
cosmological constant $\Lambda$, which is formally treated as an 
effective fluid with stress-energy tensor $T_{ab}=-\Lambda g_{ab}$ and 
constant $P=-\rho=-\Lambda$.} The fluid could be a real or an effective 
one (e.g., the effective fluid equivalent of a Brans-Dicke scalar field 
in scalar-tensor cosmology, see below).

The covariant conservation of this stress-energy tensor, $\nabla^b  
T_{ab}=0$, gives 
\begin{eqnarray}
&& u_a \left[ u^c \nabla_c \left(P+\rho \right) \right] 
+\left(P+\rho\right) 
u^c \nabla_c u_a +\left(P+\rho\right) u_a \nabla^c u_c +\nabla_a P 
 =0 \,.\label{covcons}
\end{eqnarray}
The fluid  four-velocity $u^a$ determines the $3+1$ splitting of 
spacetime into the time and space ``seen'' by observers comoving with the 
fluid. 
Their 3-space has the Riemannian metric $h_{ab} \equiv g_{ab}+u_a u_b$, 
while the operator  ${h^a}_b$ (with $h_{ab} u^a=h_{ab}u^b=0$) 
projects tensors onto this 3-space. Projecting Eq.~(\ref{covcons}) yields 
\be
{h^a}_c \nabla_a P + \left( P+\rho \right) h_{cb} a^b =0 \,,\label{cons2}
\ee
where $a^a \equiv u^b \nabla_b u^a$ is the four-acceleration of the fluid 
elements. When $P$ is constant, Eq.~(\ref{cons2}) reduces to the affinely 
parametrized geodesic equation $a^b = {h^b}_a a^a=  u^c \nabla_c u^b = 0$. 
If $\nabla_aP\neq 0$, instead, the fluid elements deviate from geodesics 
with acceleration $a_a=-\frac{ {h_a}^b \nabla_bP}{P+\rho}$.  
However, if $\nabla^a P$ points along the time direction $u^a$ of the 
comoving 
observers (parallel force), projecting it onto the 3-space orthogonal to 
$u^a$ eliminates 
it 
completely from the equation of motion of the fluid elements.


If dark matter self-interacts, it can originate negative bulk stresses, 
which have the effect of causing acceleration of the cosmic expansion and 
have been investigated as a potential explanation for the present-day 
acceleration of the cosmic expansion discovered in 1998 with Type~Ia 
supernovae \cite{Zimdahl:1996cg,Zimdahl:1997qe, 
Zimdahl:1998rx,Zimdahl:1998zq, Zimdahl:2000zm}. The self-interaction 
mechanism effectively causes forces on the dark matter fluid that are 
parallel to their worldlines, called ``cosmic antifriction'' 
\cite{Zimdahl:1996cg,Zimdahl:1997qe, Zimdahl:1998rx,Zimdahl:1998zq, 
Zimdahl:2000zm}.

\subsection{Newtonian analogue}

It is useful to resort to a Newtonian analogy to visualize the 
corresponding relativistic situation. Let us  consider a particle of 
(constant) mass $m$ subject to a friction-like force that remains 
parallel to its velocity $\vec{v}$:
\be
  \frac{d}{dt} \left( m \vec{v} \right) = \vec{F}= \alpha(t)\, 
\vec{v} \,,
\ee
where  $ \alpha(t) $ is a function of time. We want to find a new 
time parameter $\tau$ such that the new  variable $ 
\vec{u} \equiv d\vec{x} / d\tau $ satisfies $d\vec{u}/d\tau =d^2 
\vec{x}/d\tau^2=0$, i.e., the parallel force $\vec{F}=\alpha(t)\, \vec{v}$ 
is removed by the reparametrization. This problem has  
a solution: in fact, using 
\be
\vec{u} = \frac{d\vec{x}}{dt}\, \frac{dt}{d\tau} = \vec{v} 
\, \frac{dt}{d\tau} \,,
\ee
we have $\vec{v} = \vec{u} \, d\tau/dt $ and 
\begin{eqnarray}
\vec{F} &=& \frac{d}{dt} \left( m \vec{v} \right) = 
\frac{d}{dt} \left( m \vec{u} \, \frac{d\tau}{dt} \right)
= m \frac{d}{dt} \left( \vec{u}\, \frac{d\tau}{dt} \right) 
m \left( \frac{d\tau}{dt}\, \frac{d\vec{u}}{dt} + \vec{u} 
\frac{d^2\tau}{dt^2} \right) \,.
\end{eqnarray}
Using $ \frac{d\vec{u}}{dt} = \frac{d\vec{u}}{d\tau} \, \frac{d\tau}{dt} 
$, we obtain
\be
\vec{F} = m \left[ \left( \frac{d\tau}{dt} \right)^2 
\frac{d\vec{u}}{d\tau} + \vec{u} \, \frac{d^2\tau}{dt^2} \right] 
\ee
and equating this expression to the parallel force $ \alpha(t) \, 
\vec{v} $ yields 
\be
m \left[ \left( \frac{d\tau}{dt} \right)^2 \frac{d\vec{u} }{d\tau} + 
\vec{u} \, \frac{d^2\tau}{dt^2} \right]  
= \alpha(t) \, \vec{u} \,  \frac{d\tau}{dt} \,,
\ee
or
\be
\vec{u} \left( m \, \frac{d^2\tau}{dt^2}  - \alpha(t)\, \frac{d\tau}{dt} 
\right) 
+ m \left( \frac{d\tau}{dt} \right)^2 \frac{d\vec{u}}{d\tau} = 0 \,.
\ee  
Therefore, we must impose the ODE
\begin{equation} \label{eq:tau_eom}
 m \frac{d^2\tau}{dt^2}  
- \alpha(t) \, \frac{d\tau}{dt}  = 0 \,,
\end{equation} 
which gives the general solution
\be
\tau (t) = C_1 \int \exp \left( \frac{1}{m} \, \int \alpha(t')\,dt' \right) 
dt + C_2
\ee 
to the problem of eliminating the force $\vec{F}= \alpha(t) \, 
\vec{v} $ (here $C_{1,2} $ are integration constants). 

In the familiar situation in which $\alpha$ is  a negative  constant, the 
solution reduces to 
\be
\tau (t) = \bar{C}_1 \, \mbox{e}^{ -\,  \frac{ \mid \alpha \mid }{m} 
\, t }  + C_2  \,.
\ee
If a second force not parallel to $\vec{v}$ is present (for 
example, the restoring force in the damped harmonic oscillator), the time 
reparametrization generally turns it into a more complicated form.

\section{Einstein frame scalar-tensor gravity}
\label{sec:3}
\setcounter{equation}{0}

Let us turn now to  scalar-tensor gravity described by the  Jordan frame 
action 
\cite{Brans:1961sx}  
\be
S_\mathrm{ST} 
=\frac{1}{16\pi} \, \int d^4 x \, \sqrt{-g} \left[ \phi R 
-\frac{\omega}{\phi} \, g^{cd} \nabla_c\phi \, \nabla_d\phi -V( \phi) 
\right]  + S_\mathrm{(m)} \,, \label{1} 
 \ee
where  $ S_\mathrm{(m)}=\int d^4 x \, \sqrt{-g} \, {\cal L}^{(m)} $  
is the matter action, $\phi>0$ is the Brans-Dicke 
scalar field, and $\omega > -3/2$ is the ``Brans-Dicke coupling'' 
\cite{{Brans:1961sx}}. The stress energy tensor of matter is covariantly 
conserved,  $\nabla^{b} \,  T_{ab}^{(m)} =0 $ and test particles follow 
geodesics of the metric $g_{ab}$, which is deduced from the covariant 
conservation of a dust fluid as in GR \cite{Waldbook}.  

The  conformal rescaling of the metric and the scalar field redefinition 
\be 
g_{ab}\rightarrow \tilde{g}_{ab}=\phi \, g_{ab} \,, \quad\quad 
d \tilde{\phi}= \sqrt{ \frac{2\omega+3}{16\pi G} } \, 
\frac{ d\phi }{\phi} 
\ee
bring the scalar-tensor  action~(\ref{1}) into its 
Einstein frame form \cite{Dicke:1961gz} 
\begin{eqnarray} 
S_\mathrm{ST} & =& \int d^4 x \, \Bigg\{ 
\sqrt{ -\tilde{g}} \left[\frac{ 
\tilde{R}}{16\pi G} -\frac{1}{2} \, \tilde{g}^{ab}
 \tilde{\nabla}_a\tilde{\phi} 
\tilde{\nabla}_b\tilde{\phi} -U\left(\tilde{\phi} 
\right) \right] 
\nonumber \\ 
& \, &  + \, \mbox{e}^{ -8\sqrt{ \frac{\pi }{2\omega +3} 
} \,\, 
\tilde{\phi} } {\cal L}^{(m)} 
\left[ \tilde{g} \right] \Bigg\} 
\,,\label{47} 
\end{eqnarray} 
where a tilde denotes Einstein frame quantities, $\tilde{\nabla}_a$ is the 
covariant derivative of the rescaled metric $\tilde{g}_{ab}$, and 
\cite{Dicke:1961gz,mybook1}  
\be \label{47bis}
U\left( \tilde{\phi} \right) = V\left[ \phi \left( \tilde{\phi} 
\right)  \right] 
\exp \left( -8 \sqrt{\frac{\pi G}{2\omega+3} } \, \tilde{\phi} \right) \,.
\ee 
The matter Lagrangian density is now multiplied 
by the factor $ \exp \left( -8 \sqrt{\frac{\pi G}{2\omega+3} } \right) 
$ and the scalar field $\tilde{\phi}$  couples explicitly to 
matter. As a result, the units of time, length, and mass vary with $\phi$ 
and, therefore, with the spacetime position, as do particle masses 
\cite{Dicke:1961gz}. This 
fact is in principle  
immaterial since 
in an experiment one always measures {\it ratios} of quantities to their 
units and, since particle masses scale in the same way as their units, 
their ratios remain constant \cite{Dicke:1961gz}. The  
stress-energy tensor of any non-conformal matter species  fails to be 
covariantly conserved in the Einstein frame 
\cite{Waldbook,Dicke:1961gz,mybook1}: the ``new''   
$\tilde{T}_{ab}^{(m)} $ satisfies \cite{Dicke:1961gz,mybook1}  
\be \label{43} 
\tilde{\nabla}^{b} \, 
\tilde{T}_{ab}^{(m)} =-\,\frac{d}{d\phi} \left[ \ln \Omega (\phi) \right] 
\, \tilde{T}^{(m)} \, \tilde{\nabla}_a \phi 
\,.
\ee
Timelike geodesics of  $g_{ab}$ are not mapped into geodesics  of 
$\tilde{g}_{ab}$ by the conformal 
transformation because the force proportional to $ 
\tilde{\nabla}^a \phi$ makes them depart from geodesic curves. The 
Einstein frame action~(\ref{47})  yields 
\be \label{43bis} 
\tilde{T}_{ab}^{(m)}=\frac{-2}{\sqrt{ -\tilde{g} } } \, 
\frac{ \delta \left( \sqrt{-\tilde{g}} \, \, {\cal L}^{(m)} \right) 
}{\delta \tilde{g}^{ab} } 
\ee 
while,  under the conformal rescaling $g_{ab} \to 
\tilde{g}_{ab}=\Omega^2 \,  g_{ab}$, $T_{ab}^{(m)}$ 
scales according to 
\be 
\tilde{T}^{ab}_{(m)}=\Omega^s \, \, T^{ab}_{(m)} \,, \quad\quad 
\tilde{T}_{ab}^{(m)}=\Omega^{s+4} \,\, T_{ab}^{(m)} \,, 
\ee 
where $s$ is an appropriate conformal weight. The Jordan frame 
covariant conservation equation $\nabla^b \, T_{ab}^{(m)}=0$ becomes  
\begin{eqnarray} 
\tilde{\nabla}_a \left( 
\Omega^s \, T^{ab}_{(m)} \right) &=&\Omega^s \, \nabla_a T^{ab}_{(m)} 
+\left( s+6 \right) \Omega^{s-1} \, T^{ab}_{(m)}\nabla_a \Omega 
\nonumber\\
&&\nonumber\\
&\, & - \Omega^{s-1} g^{ab} \, T^{(m)} \nabla_a \Omega  \label{x1}
\end{eqnarray} 
in the Einstein frame \cite{Waldbook}. 
The choice  $s=-6$ yields \cite{Waldbook} yields
\be \label{x3} 
\tilde{T}^{(m)} \equiv 
\tilde{g}^{ab} \, \tilde{T}_{ab}^{(m)}= \Omega^{-4} \,\, T^{(m)} 
\ee 
and Eq.~(\ref{x1}) becomes  
\be \label{x3bis} 
\tilde{\nabla}_a 
\tilde{T}^{ab}_{(m)} =- \tilde{T}^{(m)}\, \tilde{g}^{ab} \, 
\tilde{\nabla}_a \left( \ln \Omega \right)  \, , 
\ee 
while
\be
\tilde{T}_{ab}^{(m)} = \Omega^{-2} \, T_{ab}^{(m)} \,, \quad\quad 
 \widetilde{ { {T_{a}}^{b}}^{(m)} } = \Omega^{-4} \, {{T_a}^b}^{(m)} 
\,,\quad\quad  
 {\tilde{T}}^{ab} = \Omega^{-6} \, {T^{ab}}^{(m)} \,, 
\label{43ter} 
\ee  
and 
\be \label{43quater} 
\tilde{T}^{(m)}= 
\Omega^{-4} \, T^{(m)} \,. 
\ee

For scalar-tensor gravity  with $\Omega=\sqrt{\phi}$ 
\cite{Wagoner:1970vr,mybook1},  
\be \label{x4} 
\tilde{\nabla}_a \tilde{T}^{ab}_{(m)} =- \frac{1}{2\phi} \,\, 
\tilde{T}^{(m)}
 \, \tilde{\nabla}^b \phi = - \sqrt{ \frac{4\pi }{2\omega
+3} } \, \, \tilde{T}^{(m)} \,\, \tilde{\nabla}^b \tilde{\phi} \,, 
\ee 
from which the modified ``quasi-geodesic'' equation can be derived. In 
fact,  a dust fluid
with zero pressure in the Einstein frame satisfies
\be
\tilde{u}_a \, \tilde{u}_b \, \tilde{\nabla}^b \tilde{\rho}^{(m)}
+\tilde{\rho}^{(m)} \, \tilde{u}_a \, \tilde{\nabla}^b \tilde{u}_b 
+\tilde{\rho}^{(m)} \, \tilde{u}_c \,\tilde{\nabla}^c \, \tilde{u}_a 
 - \sqrt{ \frac{4\pi G}{2\omega +3} } \, \,
\tilde{\rho}^{(m)} \, \tilde{\nabla}_a \tilde{\phi} =0  \label{x7} 
\ee
or, using explicitly the proper time $\tau$ along the 
fluid  worldlines, 
\be
 \tilde{u}_a \left(
\frac{d\tilde{\rho}^{(m)} }{d\tau}+\tilde{\rho}^{(m)} \, \tilde{\nabla}^c 
\tilde{u}_c \right) +\tilde{\rho}^{(m)} \left( \frac{ d\tilde{u}_a}{d\tau} 
-\, \sqrt{ \frac{4\pi G}{2\omega+3}} \, \tilde{\nabla}_a\phi \right) 
 =0 \,, \label{x8}
\ee 
equivalent to 
\be \label{x9} 
\frac{d\tilde{\rho}^{(m)} 
}{d\tau} + \tilde{\rho}^{(m)} \, \tilde{\nabla}^c \tilde{u}_c =0 
\ee 
and 
\be \label{x10} 
\frac{d\tilde{u}^a }{d\tau} =\sqrt{ \frac{4\pi 
G}{2\omega+3}} \, \, \tilde{\nabla}^a \tilde{\phi} \,. 
\ee 
With a procedure analogous to the GR case \cite{Waldbook}, the Einstein 
frame ``quasi-geodesic'' equation can be derived\footnote{This equation  
also appears in  the 
low-energy limit of string theories, but the coupling of the string 
dilaton to Standard Model particles is usually 
not  universal,  violating the Weak Equivalence 
Principle  
\cite{Taylor:1988nw,damour1994string,Gasperini:1999ne}.}  
\cite{Wagoner:1970vr,Cho:1992yj,Cho:1994qv}: 
\be \frac{d^2 x^a}{d\tau^2} 
+\tilde{\Gamma}_{bc}^a \, \frac{d x^b}{d\tau} \, \frac{d x^c}{d\tau} = 
\sqrt{ \frac{4\pi G}{2\omega + 3}} \, \, \tilde{\nabla}^a \tilde{\phi} 
\,.\label{x11} 
\ee 

Along the particle trajectory, the spacetime dependence of the 
scalar $\phi$ becomes the dependence from the parameter (e.g., proper 
time) along this curve, $\phi=\phi(\tau)$ and the  four-force is parallel 
to to its tangent,  
\be 
\frac{d^2 x^a}{d\tau^2} +\tilde{\Gamma}_{bc}^a \, \frac{d x^b}{d\tau} \, 
\frac{d x^c}{d\tau} = \sqrt{ \frac{4\pi G}{2\omega + 3}} \, 
\frac{d\tilde{\phi}}{d\tau} \, \tilde{u}^a \,.
\ee 

This parallel four-force has been correctly interpreted as a variation of 
the particle mass along its trajectory, which forces the particle away 
from a geodesic and can be eliminated by reparametrizing the worldline 
with an affine parameter different from the proper time $\tau$. The 
variation of a particle mass causes a similar effect in Newtonian 
mechanics, as discussed next.

\subsection{Newtonian analogue}

Again, a Newtonian analogy helps understanding the relativistic situation. 
For a particle of time-varying mass $m(t)$ and momentum $\vec{p} = 
m \vec{v}$ subject to zero external forces, Newton's second law 
$\vec{F}=d\vec{p}/dt $ gives 
\be
\dot{m} \vec{v} + m\vec{a}=0 \,,
\ee
where an overdot denotes differentiation with respect to the time 
$t$ and  $\vec{a}$ is the acceleration, which is 
{\em parallel to the velocity} $\vec{v}$, 
\be
 \vec{ a} = - \frac{\dot{m}}{ m} \, \vec{v} \,.
\ee
If the mass increases, the particle decelerates while it gains mass, an  
effect similar to that of a friction force, but there is no such 
force. If instead the mass decreases, the particle  accelerates, similar 
to an antifriction. This situation makes it easier to understand the 
``cosmic antifriction'' scenarios of 
Refs.~\cite{Zimdahl:1996cg,Zimdahl:1997qe, Zimdahl:1998rx,Zimdahl:1998zq, 
Zimdahl:2000zm} for self-interacting dark matter, in which the mass of 
dark matter particles varies in time.

The Lagrangian 
\be 
L=\frac{m}{2} \, \vec{v}^{\,2} =T
\ee
coincides with the kinetic energy $T$ and with the Hamiltonian 
\be
{\cal H}= \sum_i p_i v^i -L = 2T-T=T 
\ee
and does depend on time ($ d{\cal H}/dt= -\frac{ \dot{m}}{2} \,  v^2  
=- \frac{m^2}{2\dot{m}} \, a^2$). 

The particle is accelerated due to the fact  that its mass changes with 
time (an example is a rocket, which initially  has most of its mass in 
fuel and burns it in flight). This acceleration is unintuitive only if one 
thinks of the more popular version of Newton's second law 
$\vec{F}=m\vec{a}$, which is only correct for constant mass $m$: the 
correct formulation of Newton's second law is $\vec{F}=d\vec{p}/dt$. 
Similarly, in relativity one should think of the generalization of 
Newton's second law as $F^a=dp^a/d\tau$, not as  
$F^a=ma^a$. 

In Newtonian mechanics, the acceleration parallel to the trajectory can be 
formally eliminated by redefining the time.  Look for a new time parameter 
$\bar{t}(t)$ such that $ d^2 
\vec{x}/d\bar{t}^2=0$. We have 
\be
\frac{d\vec{x}}{d\bar{t}} =   \frac{dt}{d\bar{t}} \,
\frac{d\vec{x}}{dt} \,, \quad\quad  
\frac{d^2\vec{x}}{d\bar{t}^2} =  
\vec{a} \left( \frac{dt}{d\bar{t}}\right)^2  
+ \vec{v} \, \frac{d^2t}{d\bar{t}^2} 
\ee
and, using $\vec{a}=- \frac{\dot{m}}{m} \, \vec{v}$, 
\be
\frac{d^2\vec{x}}{d\bar{t}^2} =  \left[ -\frac{1}{m} \, 
\frac{dm}{dt} \left( \frac{dt}{d\bar{t}} \right)^2  
+\frac{d^2 t}{d\bar{t}^2} \right] \vec{v} \,,
\ee
which is (anti-)parallel to $\vec{v}$. 
Setting this quantity to zero  yields
\be
\frac{d^2 t}{d\bar{t}^2} = \left( \frac{dt}{d\bar{t}}\right)^2  
\frac{\dot{m}}{m_0}= \frac{dt}{d\bar{t}} \, \frac{ d\ln (m/m_0)}{d\bar{t}} 
\,, 
\ee
where $m_0$ is a constant. Integration gives 
\be
\frac{dt}{d\bar{t}} =\frac{m( \bar{t})}{m_0} 
\ee
or
\be
\frac{d\bar{t}}{dt} = \frac{m_0}{m}
\ee
and 
\be
\bar{t}(t) = \int  \frac{m_0}{m(t)}  dt \,.
\ee
Of course, time $t$ is absolute in Newtonian mechanics and the new time 
parameter $\bar{t}$ does not carry any physical meaning.  This situation 
is analogous to that of test particles in Einstein frame scalar-tensor 
gravity, where $m=m(\tau)$ along the trajectory and the four-acceleration 
parallel to the four-velocity can be eliminated by a reparametrization, at 
the price of losing the proper time.

For massive test particles in Einstein frame scalar-tensor gravity, the 
mass depends on the position along the trajectory through its 
$\phi$-dependence (therefore on the proper time, or other parameter 
labelling the position along this trajectory). The interpretation of the 
deviation from a geodesic as due to the varying mass is indeed 
appropriate.


 \section{Conclusions} 
\label{sec:6} 

In relativistic physics, forces parallel to particle worldlines appear in 
situations of high physical relevance, such as perfect fluids with 
pressure in cosmology and in the motion of test particles in Einstein 
frame scalar-tensor gravity. Yet, contrary to normal forces that have been 
discussed explicitly \cite{Price:2004ns}, parallel forces do not seem to 
have received attention. We have provided Newtonian analogies for both 
situations mentioned above to gain physical intuition. In FLRW cosmology 
(except for dust), a pressure gradient 
points in the comoving time direction, that is, parallel to the trajectory 
and the comoving time is not an affine parameter. For test particles in 
Einstein frame scalar-tensor gravity, the deviation from geodesic curves 
can legitimately be attributed to a variation of the particle mass with 
the scalar 
$\phi$, therefore, with the position along the trajectory. In both cases,  
the parallel force can be eliminated by a reparametrization, but at the 
price of losing the proper time as a parameter. The same can be 
said for the Newtonian situation, where giving up absolute time appears 
even more artificial.

\section*{Declarations} 
\subsection*{Data availability} Not applicable. 


\subsection*{Funding}
This work is supported, in part,  by the Natural Sciences \& Engineering 
Research Council of Canada (grant No. 2023-03234 to V.F.).


\bibliography{sn-bibliography}


\end{document}